\documentclass{pas}

\usepackage{multirow}

\usepackage{comment}
\usepackage{stfloats}

\begin{document}

\lefttitle{Publications of the Astronomical Society of Australia}
\righttitle{Cambridge Author}

\jnlPage{1}{4}
\jnlDoiYr{2021}
\doival{10.1017/pasa.xxxx.xx}

\articletitt{Research Paper}




\title{GBD-DART - I: Gauribidanur Diamond Array Radio Telescope: A Novel Configuration of LPDAs for Low-Frequency Observation of Pulsars and Radio Transients}

\title{GBD-DART - I: Diamond Array of LPDAs for Low-Frequency Pulsar and Radio Transient observations at Gauribidanur }

\title{GBD-DART-I : LPDAs in a diamond configuration to observe pulsars and transients between 130 to 350 MHz at Gauribidanur }

\title{GBD-DART-I : Pulsars and transient source observation between 130 MHz and 350 MHz at Gauribidanur }







\author{\gn{Arul Pandian B}$^{1,2}$, \gn{Joydeep Bagchi}$^{1}$, \gn{Prabu Thiagaraj}$^{2}$,   \gn{K.B.Raghavendra Rao}$^{2}$, \gn{Vinutha Chandrashekar}$^{2}$, \gn{R Abhishek}$^{2}$, \gn{Arasi Sathyamurthy}$^{2}$, \gn{Sandhya}$^{2}$, \gn{Sahana Bhattramakki}$^{2}$, \gn{Kasturi S}$^{2}$, \gn{Shiv Sethi}$^{2}$}

\affil{$^1$Department of Physics and Electronics, CHRIST (Deemed to be University), Bangalore, India., $^2$Raman Research Institute, Bangalore 560080, India. }

\corresp{Arul Pandian B, Email: arulpandian022@gmail.com}

\citeauth{Author1 C and Author2 C, an open-source python tool for simulations of source recovery and completeness in galaxy surveys. {\it Publications of the Astronomical Society of Australia} {\bf 00}, 1--12. https://doi.org/10.1017/pasa.xxxx.xx}

\history{(Received xx xx xxxx; revised xx xx xxxx; accepted xx xx xxxx)}

\begin{abstract}

 Gauribidanur Diamond Array Radio Telescope (GBD-DART) is a new small LPDA antenna array consisting of 64 short dipoles and associated receivers that has been custom developed and deployed at the Gauribidanur observatory (13.604 N, 77.427 E) to study bright Pulsars and Solar transients in the frequency range of 130-350 MHz. 
The LPDAs are arranged in a checkerboard layout, with opposite pairs combined to enable dual-polarised operation.  A diamond-shaped (tilted square) array configuration was chosen to achieve high sidelobe suppression in the East-West and North-South directions.  The tile measures 5.9 meters by 5.9 meters, with diagonals along both the North-South and East-West directions, each measuring about 8.4 meters.  The LPDA array with one diamond-shaped tile has been fully commissioned and is operating in transit-observing mode, successfully detecting strong pulsars and solar flares over the last seven months.  The present digital backend restricts the instantaneous bandwidth for observations to 16 MHz.  The array operations are streamlined to facilitate remote operations.  Apart from investigating Pulsar and Solar phenomena at low radio frequencies in selected sources, this work aims to provide a training platform for radio astronomy through simple-to-construct, low-cost radio telescopes.  In this paper, we present details of the array, including antenna and array response studies, brief descriptions of front-end and backend instrumentation,  and illustrative results from observations of both pulsars and solar flares.  It will also provide brief details of future upgrade plans, particularly for the tiles and digital backend, to facilitate the observation of additional sources.

\end{abstract}


\maketitle

\section{Introduction}

Our understanding of the Universe has significantly expanded due to radio astronomy, which has changed our perspectives. There are several important benefits to low-frequency radio observation. Numerous celestial objects are coherent continuum sources, such as pulsars, which emit synchrotron radio with a steep spectrum that intensifies substantially at low frequencies. The desire to investigate the distant Universe using the highly redshifted 21-cm line of neutral hydrogen (HI) at the epoch of reionisation \citep{morales2004toward, benson2006epoch}, cosmic microwave background radiation \citep{penzias1965measurement}, the study of the dark halo with the galactic rotation curve \citep{sofue2020rotation}, The Sun  \citep{ramesh1998gauribidanur}, neutron stars and pulsars \citep{hewish1969observation}, an understanding of the interstellar medium probogation effect with fast radio bursts \citep{petroff2019fast}, the frequency-dependent study by plasma dispersion \citep{cordes2016frequency}, magnetic field strength with faraday rotation \citep{lyne1989pulsar}. Low-frequency radio observations of pulsar emissions \citep{singha2021evidence} and timing measurements \citep{backer1986pulsar, tarafdar2022indian} are crucial for comprehending and characterising the effects of the interstellar medium. These observations help investigate phenomena such as chromatic dispersion, scintillation, and pulsar emission spectra. This information helps determine the timing of pulsar signals, thereby improving advanced investigations of gravitational wave detection. \citep{srivastava2023noise, bhat2018observations}.

Several large and modern aperture array radio telescopes, such as the Low-Frequency Array (LOFAR) \citep{van2013lofar}, the Long Wavelength Array (LWA) \citep{ellingson2009long}, the Murchison Widefield Array (MWA) \citep{tingay2013murchison}, the Canadian Hydrogen Intensity Mapping Experiment (CHIME) \citep{amiri2018chime}, the Hydrogen Epoch of Reionisation Array (HERA) \citep{deboer2017hydrogen}, and the upcoming Square Kilometre Array (SKA) \citep{dewdney2009square} designed to probe the low/mid radio frequency universe. The radio observatory at Gauribidanur facilitated pulsar observations at 34.5 MHz using a fat-dipole array \citep{sastry1989gauribidanur,deshpande1992pulsar,maan2015discovery}, and between 50 and 80  MHz using an LPDA array \citep{bane2022prototype}. 

 The new LPDA antenna array presented in this paper is designed to observe pulsars and solar transients at 130-350 MHz from the Gauribidanur Observatory. Additionally, we aimed to provide a cost-effective, simple radio telescope design for educational purposes, enabling a larger number of university students to engage in hands-on training in radio astronomy, particularly in pulsar and solar observations. The use of an LPDA-based aperture array in the design achieved cost-effectiveness and broadband frequency coverage \citep{isbell2003log}. Finding radio-frequency interference (RFI)-free zones and suitable array configurations to maximise the array gain required significant investigations \citep{arul2025a}, \citep{arul_pandian_b_2025_15709357}. The functionalities of the antenna, signal conditioning, and signal transport components were first verified in a two-element interferometer setup  \citep{likhit2025innovative}. Subsequently, a 64-dipole tile was commissioned at the observatory site. The 64 LPDAs of the array are phased to observe transiting sources nominally for about an hour at the zenith. The current instantaneous observing bandwidth is 16 MHz per polarisation, with the restriction imposed by the data recorder. The array observations can be remotely scheduled and monitored. Work is also in progress to implement a second tile to enhance observations.  

This paper outlines how we addressed the design challenges of implementing a radio array, its end-to-end functionality, and the results of observations of radio pulsars and solar transients. Section 2 discusses the project's motivation, design considerations, and choices made during the process. Section 3 examines the antenna responses investigated for different LPDA configurations, ranging from a single dipole to a diamond-shaped array. Section 4 introduces the signal conditioning, transport, data capture, and analysis pipeline to validate the end-to-end system in the field. Section 5 presents the results of field testing the system. Finally, section 6 provides a summary of the work, highlighting future directions and conclusions.

\section{Array Design considerations}

A significant constraint on the array design arises from pulsar observations: pulsar signals are extremely weak compared to those from the Sun, and they are point-like sources in the sky. Hence, one primary consideration for the array is to achieve good sensitivity at 150-200 MHz, as at these frequencies the pulsars are typically bright and tend to show emission turning points. A second consideration arises because pulsar signals originate from unresolvable tiny angular regions in the sky; the array's ability to measure broadband signals from pulsars over a narrow area of the sky without grating or side-lobe contamination is crucial. 

\begin{figure}[h]
    \begin{center}
    \includegraphics[width=\linewidth]{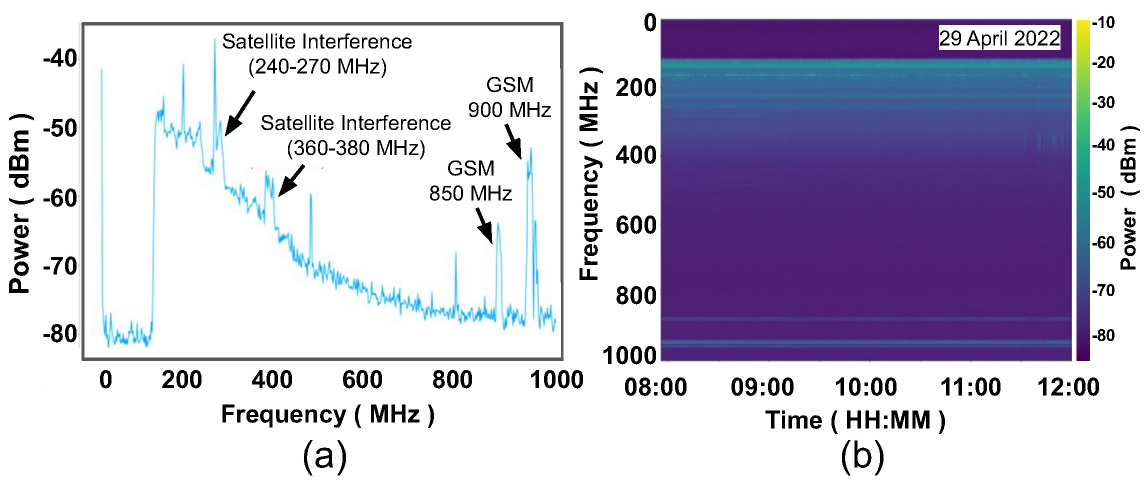}
    \end{center}
    \caption{Typical (a) spectrum and (b) spectrogram observed in the RFI measurements at the Gauribidanur observatory from 2022.  } 
    \label{fig:broadband_rfi_study}
\end{figure} 

We studied the site's RFI conditions over several days, using a custom-built receiver chain and a commercial spectrum analyser \citep{arul_pandian_b_2025_15709357}. An illustration of a typical spectrum obtained with the most numerous interferences is shown in Figures \ref{fig:broadband_rfi_study}a and \ref{fig:broadband_rfi_study}b. Strong RFI was observed between 80 MHz and 129 MHz, and moderate levels of RFI were detected between 250 MHz and 270 MHz, as illustrated in the figures. Typically, strong low-frequency (below 130 MHz) RFI saturated the measurement system's front-end amplifiers. Hence, we used a high-pass filter to eliminate low-frequency RFI bands (below 129 MHz) for the RFI study, and also set the array's lowest frequency to 130 MHz. The highest frequency for the antenna array was limited to 350 MHz due to the interference seen beyond this frequency.  

\section{\textbf{LPDAs in a Diamond Configuration} }

\begin{figure*}[t]
    \begin{center}
    \includegraphics[width=0.8\textwidth]{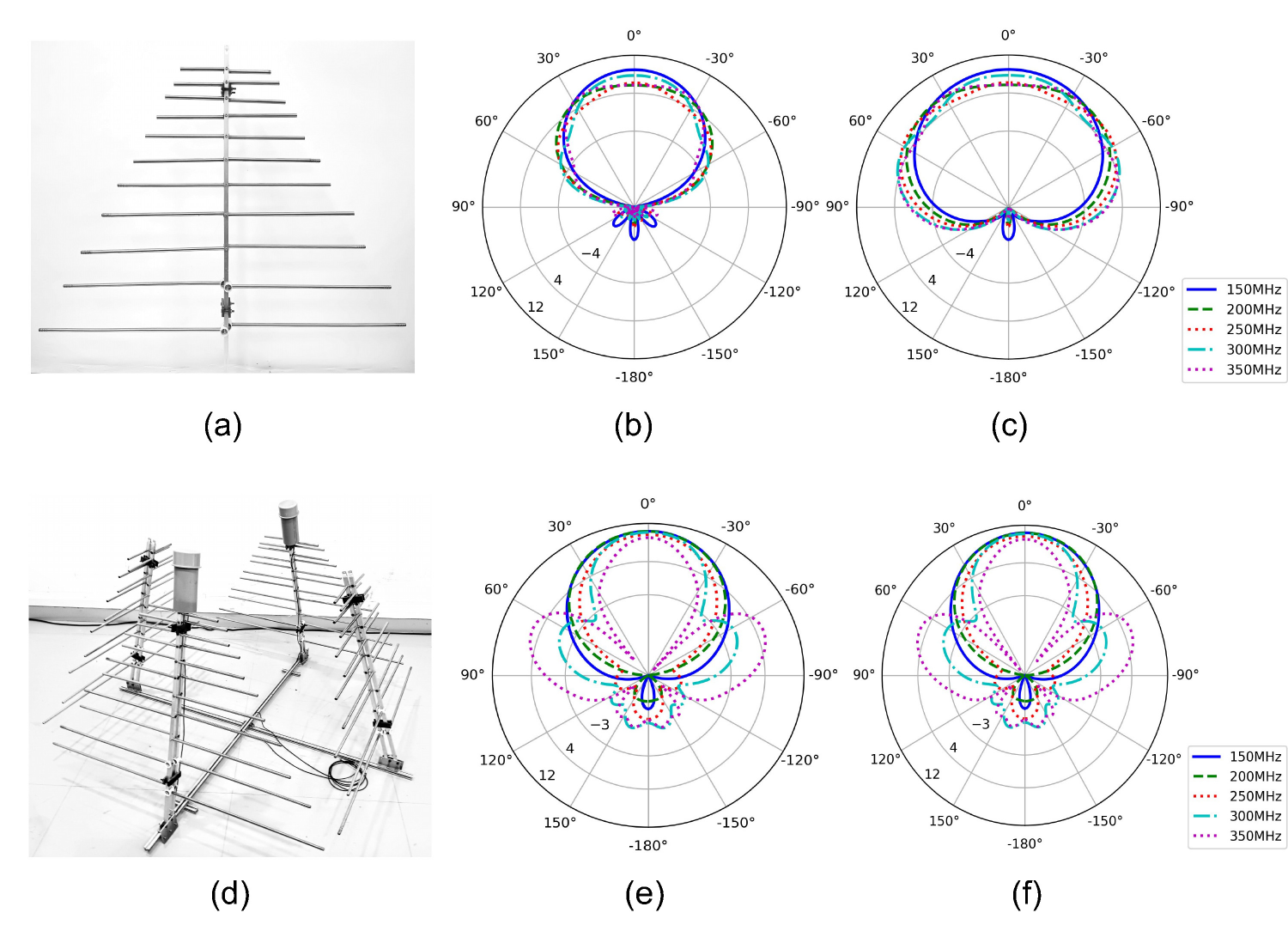}
    \end{center}
    \caption{(a) Single LPDA Antenna. The second boom with complement dipoles is joined along the side to form the 11-element LPDA. { The responses obtained at 150, 200, 250, 300 and 350 MHz are overlaid for E-plane in (b) and H-plane in (c). (d) Pyramid antenna element (e) Pyramid antenna element E-plane gain  (f) Pyramid antenna element H-plane gain .
    The simulations carried out using the }CST\textsuperscript{\tiny\textregistered} software package}. 
    \label{fig:LPDA antenna}
\end{figure*}

\begin{figure}[h!]
    \begin{center}
    \includegraphics[width=0.75\linewidth]{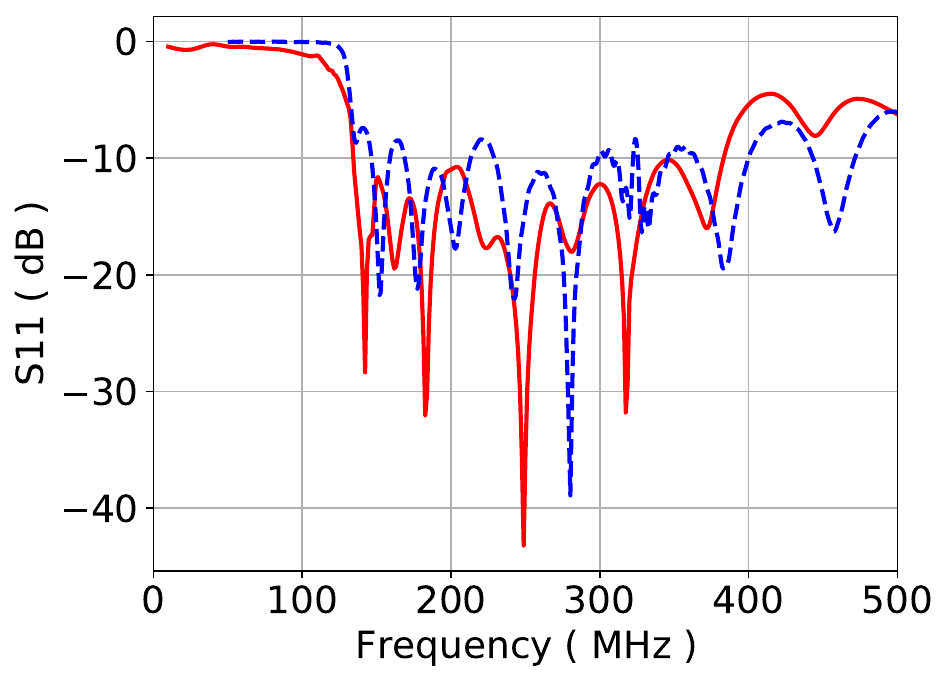}
    \end{center}
    \caption { Simulated S11 (blue dashed line) of the LPDA from CST software  is compared with measured S11 (solid red line) by N9916B microwave analyser. }
    \label{fig:S11_overlay_plot}
\end{figure}

We designed the aperture array using a set of 64 short LPDAs, which was previously custom-designed in our laboratory \citep{arul_pandian_b_2025_15709357}. The LPDA shown in Figure \ref{fig:LPDA antenna}a is an optimised version of it, mainly to operate within the 130-350 MHz frequency range, and its performance verified through beam pattern simulations shown in Figures \ref{fig:LPDA antenna}b and  \ref{fig:LPDA antenna}c, and through the S11 simulations and verifications in the field as shown in Figure \ref{fig:S11_overlay_plot},. The LPDA consisted of 11 dipole elements, each constructed from a 9 mm-diameter aluminium alloy round tube with a 1 mm wall thickness. The booms of the LPDA are fabricated from a 12 mm aluminium alloy square tube with a 1 mm wall thickness. The dipoles are connected to the boom by welding. 

 To get the orthogonal X and Y polarisations, we have examined arranging \textit{two-pairs of LPDAs} in an off-axis arrangement \citep{shankar200950,maan2013rri} as shown in Figure \ref{fig:LPDA antenna}d.  In this configuration, the voltages of opposite LPDAs are combined to form the two polarisation feeds with their electrical centre at the baseline's centre. Simulations indicate achieving a symmetric response across the broadband with a nominal ~60$^{\circ}$ field of view and ~11 dBi gain as shown in Figures \ref{fig:LPDA antenna}e and \ref{fig:LPDA antenna}f.

\begin{figure}[h!]
    \begin{center}
    \includegraphics[width=1.0\linewidth]{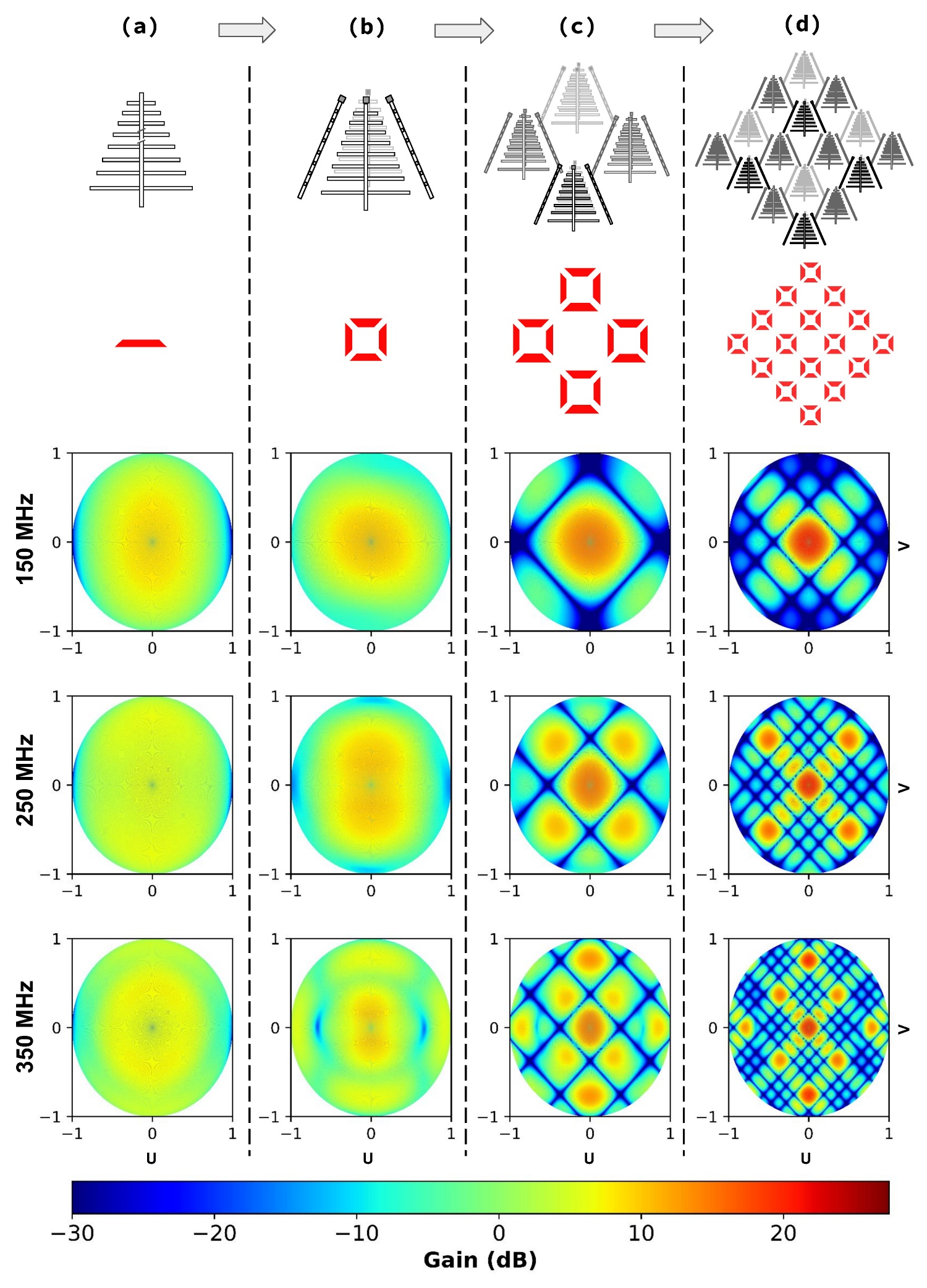}
    \end{center}
    \caption{Far-field orthographic responses of the diamond array with only single, two, eight and 32 LPDAs at 150, 250 and 350 MHz}
     \label{fig:antenna_config}
\end{figure}

 Our investigation into arrays identified a novel diamond (a 45$^{\circ}$ rotated square) configuration \citep{kraus1967radio, kraus1999electromagnetics}. A simple model of it was first studied using both \textit{Python} based and CST\textsuperscript{\tiny\textregistered} designs. We enhanced the configuration with two additional features. First, we arranged the dual-polarised LPDA pairs in a checkerboard layout \citep{kawano2016grid, kawano2017diamond, hotan2021australian} within the diamond shape. Next, we positioned the LPDA pairs at an inclined angle of 23$^{\circ}$ from the zenith ( 67$^{\circ}$  from ground) with the tilt being towards each other within the pair as is shown in \ref{fig:LPDA antenna}d. The resulting array exhibited near-flat gain across the band and significant sidelobe suppression in the East-West and North-South directions, both highly desirable. The far-field orthographic responses (obtained using CST) of the array with one, two, eight and 32 LPDAs at 150, 250 and 350 MHz are given in \ref{fig:antenna_config}.

\begin{figure}[h!]
    \begin{center}
    \includegraphics[width=1.0\linewidth]{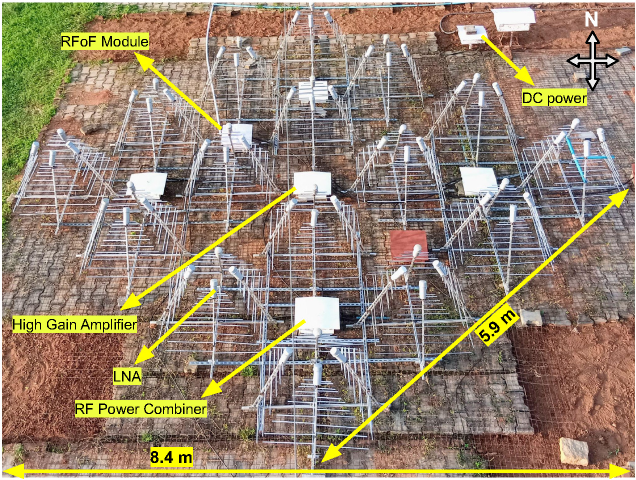}
    \end{center}
    \caption {
     Diamond antenna array deployment in the field. White boxes seen house first stage analog electronic components. A metal mesh with gap of 1/16 th wavelength of the highest operating frequency isolates the array form ground. }
     \label{fig:antenna_in_field}
\end{figure}

\section{\textbf{Tile Signal Flow}}


\subsection{\textbf{Overview}}
The array is located approximately 250 meters away from the observatory's receiver room. A comprehensive radio-frequency signal conditioning electronics: the analog font-end receiver, a fiber transmission/reception module: RFoF module, were specially developed; and an existing backend digital receiver: PDR was employed for data recording; and also a new set of software utilities for real-time data capture, data processing and archival was developed and executed as pipelines across \textit{Intel-i7} and \textit{AMD-Ryzen-9} desktops to process and store the acquired data \citep{arul_pandian_b_2025_15709357}. This section provides an overview of the array signal conditioning and data processing. A detailed discussion of it is presented in \citep{arul25pipeline}. 
 
\begin{figure*}[h]
    \begin{center}
    \includegraphics[width=0.65\textwidth]{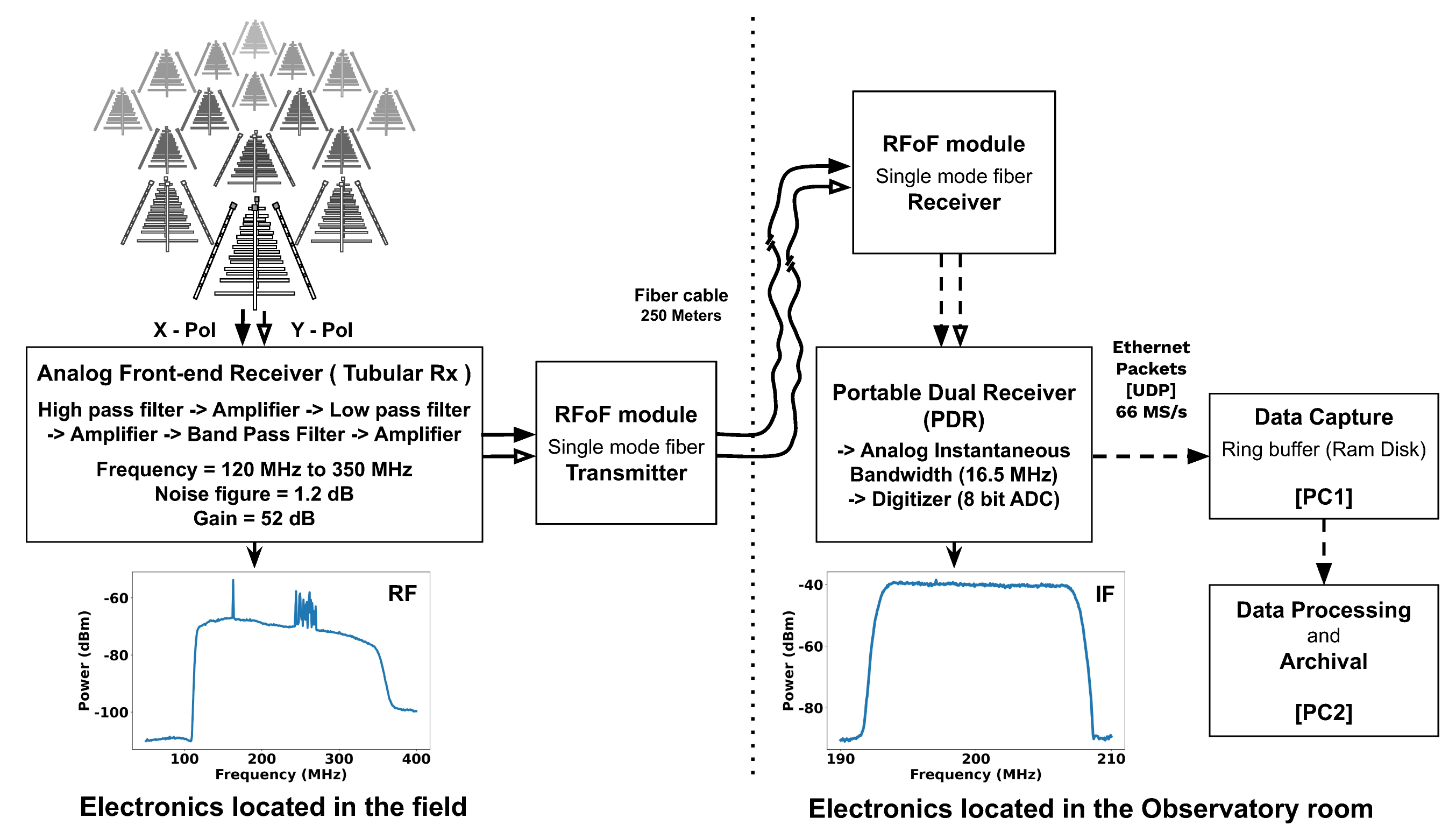}
    \end{center}
    \caption{ Array signal chain consists of 64 LPDAs in a dual polarised configuration with LNAs, voltage combiner network, high gain amplifiers with 130 to 350 MHz band-pass filters, RF to optical converters, 300 m long 1310 nm single mode optical fibers, optical to RF converters, amplifiers, 16 MHz portable analog/digital receiver, and data recording/processing/archival computers.}
    \label{fig:Signal chain GBD} 
\end{figure*} 

\subsection{\textbf{Signal Conditioning and Transport}} 
As outlined in Figure \ref{fig:Signal chain GBD}, the tile analog signal chain begins with low-noise amplifiers (LNAs) fitted at the boom feed points of the 64 LPDAs (corresponding to 32 X and 32 Y polarisation) through a BALUN arrangement. The LNAs have a 20 dB gain and a bias-tee arrangement to obtain the DC power. The amplified signal leaves the LNAs via a 3-meter-long RG-174 RF cable. The voltage signals from the X- and Y-polarisation LPDAs are combined independently using equal-length wires to produce the tile's phased-array voltage outputs for the two polarisations by combining them in two stages: eight 8-way combiners in the first stage and two 4-way combiners in the second. The first-stage combiners incorporate a bias-tee network for distributing the DC supply to the LNAs. 

 The tile's voltage outputs are further amplified by a high-gain amplifier stage in a tubular receiver (TubRx) module. It consists of multiple-stage amplifiers and filters arranged in the following sequence: a first stage amplifier, a high-pass filter allowing signals above 130 MHz, a second amplifier, a Low-pass filter to stop signals above 350 MHz, and a third amplifier. The TubRx has a noise figure of 1.2 dB and a gain of 52 dB across the 220 MHz-wide band from 130 to 350 MHz. The amplified RF signals from the TubRx are routed to an RF over Fiber (RFoF) transmitter (RFoF-Tx) module via a short RF cable. 

The RFoF-Tx module is fitted with a pre-amplifier to bias the laser diode, and it has a dynamic range of 40 dB with an operating midpoint at -40 dBm. The RFoF-Tx converts the RF signal amplitudes to a 1310 nm laser output. The transmission is by the intensity modulation of the laser output. A commercial 12-core, 250-meter-long single-mode fiber cable carries the laser signals to the observatory receiver room.  

The electronics located in the observatory room, as shown on the right side of figure \ref{fig:Signal chain GBD}, receive the optical signals from the field in an RF over fiber receiver (RFoF-Rx) module that converts the laser signal to analog RF signal. A one-meter long semi-flex cable with a shield attenuation of about 40 dB was used to carry the RF signal from the RFoF-Rx to a portable dual receiver (PDR) for subsequent band selection, digitisation, and recording of the two polarisation  signals.

\subsection{\textbf{Data-Processing}}
{ 

The portable dual receiver (PDR) is a dual-channel, 16-MHz-wideband heterodyne receiver and digitiser. The PDR can select any 16 MHz bands within the 220 MHz wide band received from the array. The PDR has dual 8-bit digitisers sampling at 33 MSPs. The sampled data from the two digitisers are timestamped, packed into a UDP packet, and sent over the Ethernet ports to a data recording computer. An onboard Virtex-5 FPGA is used in the PDR for time stamping and packetisation.

The UDP packets are first captured in the data recording computer (PC1) in a RAM-based circular buffer by using the GULP utilities \citep{Corey_Satten_2008}. The captured data fills the circular buffer at a nominal rate of 66 Mbps. The GULP writes the data in a PCAP format. A dedicated process in the PC-1 takes a copy of the data from the circular buffer, computes all correlation products via an FFT, performs a temporal average, and archives the data with timestamps. This archive is written in an HDF5 format. The archived data is used to verify the system's health and monitor the radio spectrum. For pulsar observations, the high time resolution voltage data from the ring-buffers are moved through a transient buffer to the hard disks for post processing \citep{arul25pipeline}.

\section{\textbf{Results from the Array}}

We present four different results from commissioning the array to illustrate it's end-to-end operation:
\begin{itemize}
    \item Satellite and Sun transit observations
    \item Drift observation over days
    \item Solar transient detection 
    \item Pulsar signal detection  
\end{itemize}

\subsection{\textbf{Satellite and Sun transit observations}}
\begin{figure}[h!]
    \begin{center}
    \includegraphics[width=0.65\linewidth]{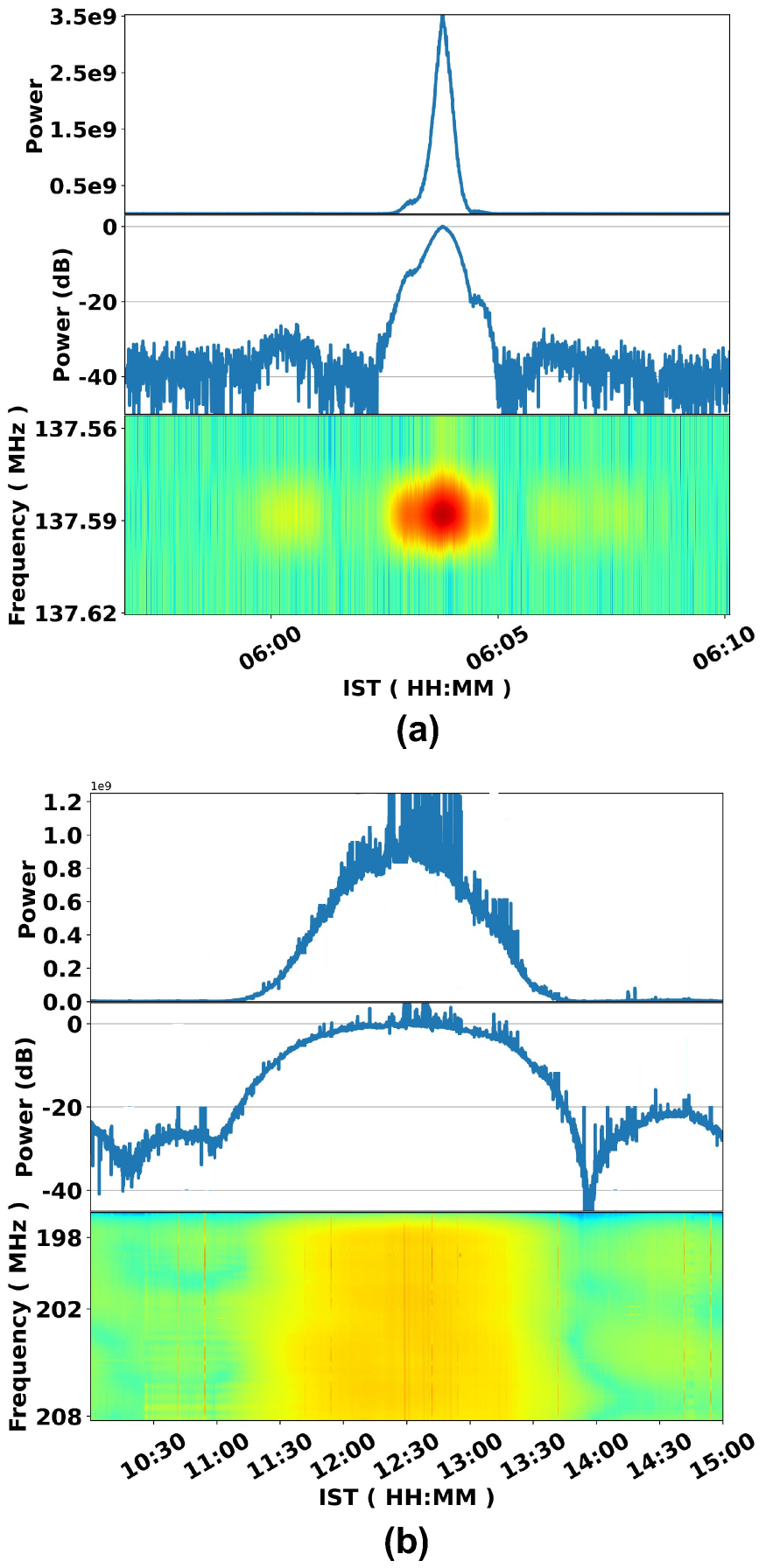}
    \end{center}
    \caption {The array's sub-group beam response study using a sub-group of eight LPDAs.  (a) Transit observation of ORBCOMM satellite pass at 137 MHz. (b) Transit observation of the Sun at 200 MHz. See also figure \ref{fig:sky_ant_mon_data_overplot} for a multiple day transit observation.} 
    \label{fig:transit_observations} 
\end{figure} 

We observed ORBCOMM satellite signals at 137 MHz by configuring the PDR mixer's local oscillator to select the lowest 16 MHz frequency band. The data were collected for a full day to cover multiple satellite passes. Figure \ref{fig:transit_observations}a illustrates a cross-section of the beam, traced during the satellite transit. A drift observation of the Sun was also made, using a band centred around 200 MHz, and the results are presented in {figure} \ref{fig:transit_observations}b. In both observations, a subgroup configuration with 8 LPDAs, as shown in Figure \ref{fig:antenna_config}, was used, and the observed side-lobe levels of around -20 dB, agree with the simulations.

\subsection{\textbf{ Drift Observation over days }}

 Drift mode observations in a sub-array configuration of Figure \ref{fig:antenna_config}c for a 16 MHz band around 200 MHz were made on multiple days to study the receiver performance over the day and against the sky temperature variations. The measured power throughout the day was compared with that expected for the sub-array's beam at the sky positions. On the 408 MHz skymap \citep{remazeilles2015improved}, a 90$^{\circ}$ declination range strip centred at +13.6$^{\circ}$ N  (instrument zenith) was convolved with the subgroup beam of Figure \ref{fig:antenna_config}c \citep{arul_pandian_b_2025_15709357}. The estimate thus-obtained is plotted as a black dot-dashed line in the upper subplot of Figure \ref{fig:sky_ant_mon_data_overplot}. The  different days measurements were normalised to the galactic plane's peak power at 18:30 hours and overlaid in various colours. The estimate matches the power observed by the array, with minor deviations. The Sun (a $\approx96000 Jy$ source)  positions shown as  the yellow dots (P1 and P2) in figure, drifts to the right during the observed epochs between 22 March 2025 and 14 April 2025.

We have also investigated the array’s responses to a few radio sources:  Cygnus-A ($8100 Jy$), Taurus-A ($1420 Jy$), Virgo-A ($970 Jy$) at 178 MHz \citep{kraus1967radio} and infer the array subgroup (Eight LPDAs) sensitivity to be around $\approx250Jy$ for 0.5 second integration with $\approx$ 2 MHz bandwidth.

\begin{figure*}[h]
    \begin{center}
    \includegraphics[width=0.9\linewidth]{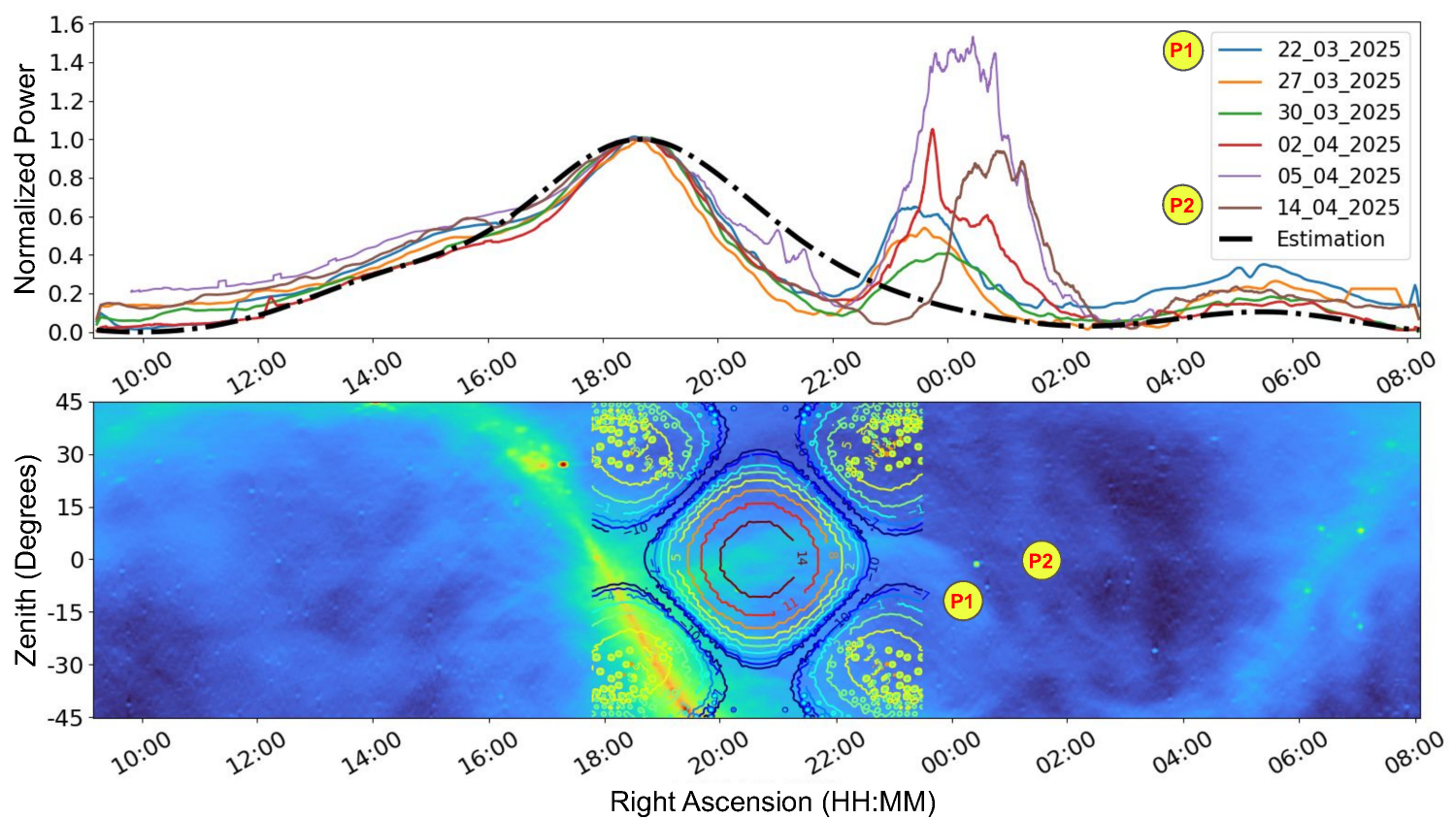}
    \end{center}
    \caption{{  The upper plot shows an overlay of results from six days' drift mode observation by the array subgroup {Figure} \ref{fig:antenna_config}c. High intensities correspond to the Galactic plane, which is visible at RA 18:30, and is obscured by the Sun between 23:00 and 02:00 Hours. 
    The plots correspond to observations made over three weeks. The lower subplot shows the radio sky visible to the array's subgroup in the  \citep{remazeilles2015improved} 408 MHz sky map. An orthographic beam pattern of the array subgroup at 200 MHz is overlaid on the sky map. This orthographic pattern was convolved with the sky map to estimate the expected total power deflections from the subgroups (shown as a black dot-dashed line in the upper plot). The total power observed was normalised to the peak power in the galactic plane at 18:30 hours. The positions of the Sun at the beginning of this set of observations on 22 March 2025 and at the end of the observation on 14 April 2025 are marked as P1 and P2, respectively, on the map.  } \color{black} }
    \label{fig:sky_ant_mon_data_overplot} 
\end{figure*} 

\subsection{\textbf{Solar Flare Detection}}

During the system commissioning period, continuous drift mode observations were conducted for system monitoring and pipeline development. Receivers recorded array data for a few hours each day during the solar transit for several months. In {figure} \ref{fig:solar_event_comparision}a, we present an intense type III Solar event (solar flare) of 27 July 2024 from the archival data captured by the array. The data have a temporal resolution of 0.5 seconds and a spectral resolution of 64 kHz. The observation was carried out at 200 MHz with a 16.5 MHz bandwidth from 9:00 AM to 3:00 PM local time. The solar event intensity across the observed band is shown on the top plot, and the dynamic spectrum obtained during the event is displayed in the bottom-left subplot. The spectrum observed during the event is given in the bottom-right plot. The observation data corresponding to the results presented are available on GitHub \citep{arul_pandian_b_2025_15709357}.  We find that our detection time and event morphologies significantly matched the same event recorded by the e-CALLISTO Solar Radio Spectrograph at the Udaipur Solar Observatory \citep{upadhyay2019solar}.  

\begin{figure}[h]
    \begin{center}
    \includegraphics[width=0.8\linewidth]{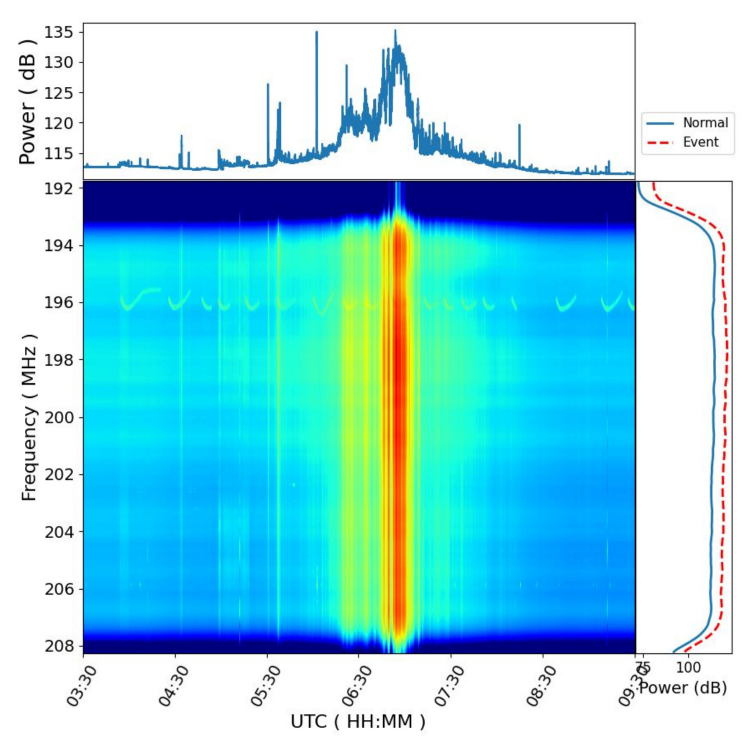}
    \end{center}
    \caption{An intense solar event recorded on 27 July 2024. }
    \label{fig:solar_event_comparision}
\end{figure} 

 The antenna power deflection during the solar event is shown in the top subplot. The right-hand subplot shows channel power as a function of frequency. The central subplot shows the dynamic spectrum over time, with frequency along the Y-axis. This spectrogram shows dynamic changes in power across channels over time, with increasing intensity from blue to red. 


\subsection{Pulsar Signal Detection}
\begin{figure}[h]
    \begin{center}
    \includegraphics[width=1.0\linewidth]{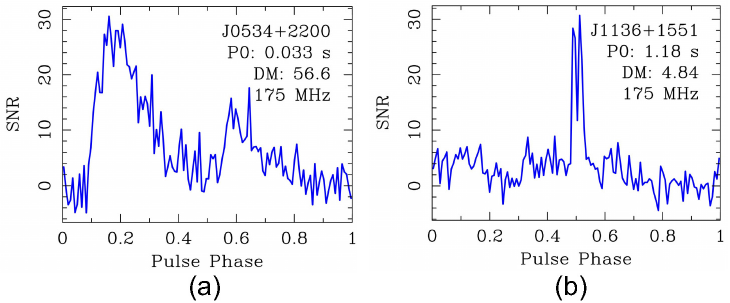}
    \end{center}
    \caption{Pulsars detected at 175 MHz with 1800 seconds integration.  (a) J0534+2200 (MJD 60945) (b) J1136+1551 (MJD 60939)  }
    \label{fig:pulsar_observation}
\end{figure} 

A primary focus of this array design was to observe a set of bright pulsars. Hence, we continually developed strategies and tools for observations and data processing \citep{arul25pipeline}. On 14th April 2025, the array had its first light, detecting pulsar B1919+21 with an SNR of 7, using only 16 LPDAs across two X-polarisation subgroups. Subsequently, we have brought up the whole array with 64 LPDAs as outlined in the paper with dual (X and Y) polarisation and successfully detected four other bright pulsars that transit across the array's nominal half-power beam between declination of +6$^{\circ}$ and +22$^{\circ}$ ({Figure} \ref{fig:sky_coverage_pulsar_obs}). A dedicated pulsar data processing pipeline was built with the standard software packages such as DSPSR \citep{van11} and PSRCHIVE \citep{hot04}. The continuous raw ADC data was recorded with the PDR system ({Figure} \ref{fig:Signal chain GBD}). The data were reduced to a time resolution of 128 $\mu$s and a spectral resolution of 64 kHz, with all stokes products in PSRFITS format \citep{hot04}. We present in {Figure} \ref{fig:pulsar_observation}, a folded total intensity pulse profile for two pulsars, J0534+2200 (CRAB pulsar) and J1136+1551. They were observed with the full array in the dual-polarised LPDA arrangement described in the paper, over a 16 MHz band centred at 175 MHz and an integration time of 1800 seconds. The intensities from the two polarisations were added during post-processing. These two pulsars represent the extremes of dispersion measures and periods in our sample of five currently detected pulsars: J0953+0755, J0837+0610, and J1921+2153, which are routinely detected by the array. The relevant findings from these additional pulsars are presented in \citep{arul25pipeline}.

\section{Future Scopes}
The array's system noise temperature of $\approx250 Jy$ for 0.5-second integration with $\approx$ 2 MHz bandwidth was estimated by observing Virgo-A. The signal strength of pulsars in our frequencies of interest typically ranges from a few milli-Jansky to a few thousand Jansky. There are 16 dual-polarised elements in the array, each with a typical gain of 10.5 dBi throughout the band, resulting in an effective aperture area of 30 $m^{2}$, an array gain of 22 dBi and a system equivalent flex density (SEFD) of 25 kJy. Thus, if we consider a 16 MHz band observation over an hour with a nominal HPBW of 15 degrees at 200 MHz, the system will be sensitive enough to detect sources with a flux density of 50 mJy.   

\begin{figure}[h]
    \begin{center}
    \includegraphics[width=\linewidth]{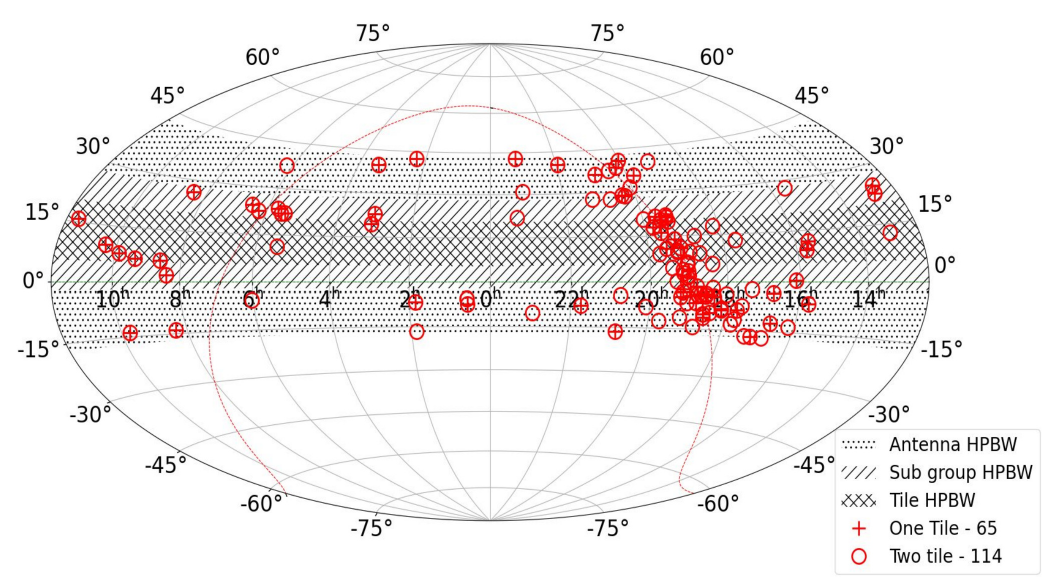}
    \end{center}
    \caption{ The sky visible to a GBD-DART tile, and detectable pulsars. The hatched region depicts the HPBW of the primary X- and Y-polarisation LPDAs. A narrower inner shaded sky area is visible for the 16 LPDA subgroup beams. The '+' symbols on the map indicate the locations of the 65 pulsars detectable with a reasonable SNR (10) in an hour of observation in a 16 MHz band around 200 MHz. The 'O symbols indicate the locations of 114 pulsars that can be detected by doubling the collecting area (two tiles). }
    \label{fig:sky_coverage_pulsar_obs}
\end{figure} 
 
Figure \ref{fig:sky_coverage_pulsar_obs} displays the regions of sky and pulsars in those regions that have flux above 50 mJy at 200 MHz, and hence are detectable for different configurations of the array. The sky areas shown by the three patches in the figure correspond to the skies that become visible in three different beamforming modes of observation. The sky region indicated by the central patch corresponds to the sky observable by a tile with its zenith beam (current mode), as determined by combining all LPDAs. The adjoining patch of the sky becomes visible to the array in groups of 8 dipoles, phased separately and combined. With this mode, additional pulsars can be observed. The outside patch area of the Sky becomes observable if the beams are formed at the individual dipole element level of a tile. The number of pulsars shown in the legend of {figure} \ref{fig:sky_coverage_pulsar_obs}b is based on the theoretical estimation \citep{lorimer2005handbook}, considering the sensitivity limits of a single tile.

\begin{figure}[h!]
    \begin{center}
    \includegraphics[width=0.9\linewidth]{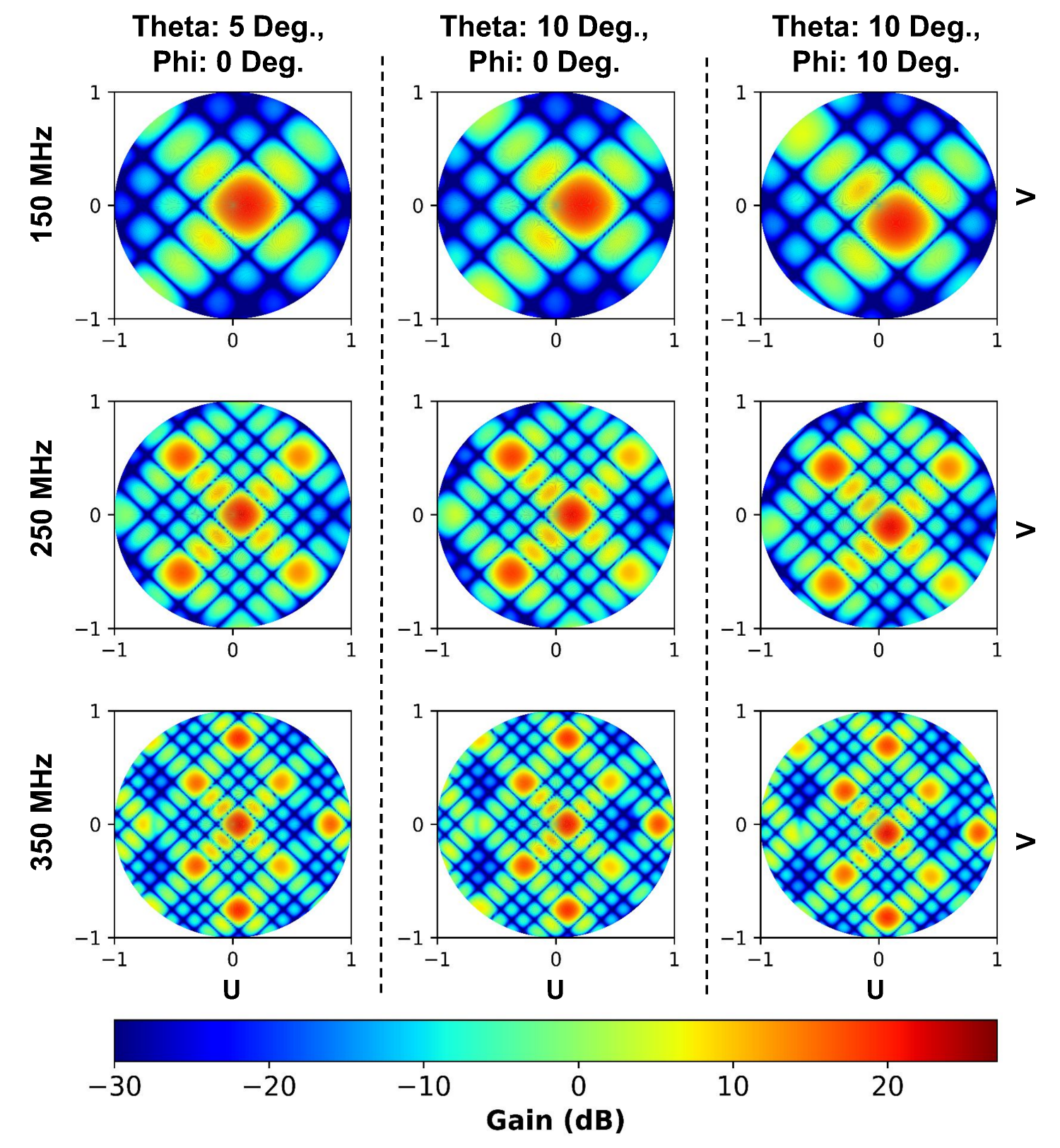}
    \caption{ Illustration of beam positions across the band with phasing.  A time-delay-based beam-forming method enables wider tracking and pointing.
    }
    \label{fig:Orthographic_plots_beamforming}
   \end{center}
\end{figure}

The array's response towards three different pointing directions at two different frequencies was simulated in CST, and the results are shown in Figure  \ref{fig:Orthographic_plots_beamforming}. To obtain the arrays' steered-beam responses, we first created the antenna array element positions in CST, exported them to a custom-developed computer code, and estimated the phase required for each primary element to steer the beam. Then, we applied these phases to the excitation ports in the CST simulation and obtained orthographic responses for each pointing set. Figure \ref{fig:Orthographic_plots_beamforming} presents the beam formation at the two extreme frequencies towards three directions in the sky.

We would also investigate forming multiple simultaneous beams for transient watch/search applications in such a case. An FFT-based beamforming option can be investigated for this array \citep{arul_pandian_b_2025_15709357}. Multiple beams are beneficial for transient searches, such as investigations into Fast Radio Bursts (FRBs).

The sensitivity can also be improved by increasing the number of tiles and enhancing the associated signal conditioning and processing. 

 Our current digital back-ends record only a 16 MHz band, and we would also consider recording and processing larger bands up to 200 MHz. These efforts would involve upgrading the digital receivers to higher-capability FPGA-based receivers, such as those recently developed in the lab \citep{girish2023progression}.

\section{\textbf{Summary and Conclusion}}

We have designed and implemented an array of log-periodic antennas (LPDA) in a diamond configuration, (measuring 5.9 meters by 5.9 meters, with its diagonals aligned along both the North-South and East-West directions) at the Gauribidanur observatory (GBD-DART). It is a small array with 64 LPDAs arranged in a checkerboard layout, forming orthogonal polarisation with dual-tilted LPDAs. This array is primarily employed to study bright Pulsars and Solar transients in the frequency range of 130-350 MHz. The diamond-shaped (tilted-square) array configuration helped suppress sidelobes. It enabled pulsar observations, often without being affected by strong signals from the Sun or other bright radio objects in the Sky, or by any low-level RFI from the horizon. All associated electronics for the array have been custom-developed and commissioned on-site. The antennas, RF front-end, fiber transmission system, back-end RF and digital systems, as well as the data-gathering and processing pipelines, are all developed in-house, and their salient features are outlined in this paper and a detailed discussions about the specialised signal processing pipeline developed to perform astronomical data analysis, particularly for pulsar data processing in \citep{arul25pipeline}. The array beam formations were studied through simulations and verified by observing satellite signals and strong celestial sources. The array also continuously collects 1 s time-resolution spectral data in a 24x7 manner, and from this archival data we detected a strong radio flare from the Sun, with the event time and morphologies verified against established observatory results. The array's ability to detect weak radio sources is demonstrated through the successful observation of five bright pulsars of varying periods, flux densities, and dispersion measures. These pulsar parameters, including profile and dispersion measure, detected by the array significantly match the standard catalogues, confirming that our frequency and timing standards are well understood and meeting the standards for such time-domain astronomy. Currently, we can observe pulsars with flux densities above 700 mJy and that transit our nominal zenith beam between +8$^{\circ}$ and +21$^{\circ}$ declination for about an hour. However, we discussed schemes to enhance the array sensitivity and detect about 10 times weaker sources. One of the primary focuses of this array building effort is to illustrate how a small radio telescope can be built to observe radio transients such as pulsars, and to leverage duplication of such designs to enhance the number of student participants both in the instrumentation and in understanding the radio pulsar observations, data analysis, and develop interest in pulsar astrophysics. This new small pulsar array is open for student training, offering hands-on experience in observations, data analysis, and signal processing for astrophysics \citep{adithya2025rass}, \citep{mahek2025rass}.

\subsection*{Disclosures}
The authors declare there are no financial interests, commercial affiliations, or other potential conflicts of interest that have influenced the objectivity of this research or the writing of this paper.

\subsection{Conflicts of Interest}
None.

\subsection*{Code, Data, and Materials Availability}
The data presented in this article are publicly available in \url{https://github.com/Arul16psp05/supplementary_materials.git} at  \url{https://doi.org/10.5281/zenodo.15709357}

\subsection*{Acknowledgments}
We acknowledge the multiple technical consultations with R. Somashekar. We also value Keerthipriya Sathish for guiding the development of the RF over fiber transmission links. We acknowledge the involvement of our colleagues at the Gauribidanur Observatory team for their support during the commissioning of the LPDA array, particularly Srinath and Janardhanan, and for their contributions to making all of the 64 LPDAs and in making the RF module enclosures at the observatory workshop and Ibrahim for the access to the RRI workshop. Additionally, we thank the undergraduate interns at Gauribidanur AES National College for their help in transporting and mounting the array of elements from the Bangalore lab to the observatory. We are also grateful to our EEG colleagues for their various forms of support and valuable conversations that aided this work. We thank the Raman Research Institute for supporting this development work. We also thank the Christ University for recognizing the research aspect of this effort. Also, we acknowledge the use of the RRI library, which provided access to the Grammarly \citep{fitria2021grammarly} tool and the publicly available QuillBot \citep{fitria2021quillbot} to correct grammar in the manuscript.

\end{document}